\documentclass [prx,twocolumn,longbibliography,superscriptaddress]{revtex4-2}

\usepackage{graphicx}
\usepackage{subfigure}
\usepackage{amsmath}
\usepackage{amsthm}
\usepackage{amssymb}
\usepackage{hyperref}
\usepackage{xcolor}
\usepackage{textcomp}
\usepackage{orcidlink}

\hypersetup{colorlinks=true, linkcolor=black, citecolor=black, urlcolor=blue}

\newcommand{\nn}{\nonumber}

\begin{document}

\title{Rotational Superradiance in a Time-Reversal Symmetry-Broken Quantum Gas inside an Optical Cavity}
 
\author{Natalia Masalaeva\,\orcidlink{0000-0002-9314-8973}}
\email[Corresponding author: ]{natalia.masalaeva@uibk.ac.at}
\affiliation{Institut f\"ur Theoretische Physik, Universit{\"a}t Innsbruck, A-6020~Innsbruck, Austria}
\affiliation{Institut f\"ur Experimentalphysik, Universit{\"a}t Innsbruck, A-6020~Innsbruck, Austria}
\author{Farokh Mivehvar\,\orcidlink{0000-0003-4776-1352}}
\email[Corresponding author: ]
{farokh.mivehvar@uibk.ac.at}
\affiliation{Institut f\"ur Theoretische Physik, Universit{\"a}t Innsbruck, A-6020~Innsbruck, Austria}

\begin{abstract}
Appearance of quantized vortices in a superfluid and a Bose--Einstein condensate (BEC) stems from their nontrivial response to \emph{broken} time-reversal symmetry (TRS). Here we show that breaking of the TRS by, for example, rotation or an external synthetic magnetic field in a transversely-driven BEC coupled to a single mode of an optical cavity modifies drastically Dicke-superradiance and self-ordering phenomena in this system. In particular, photon scattering from the pump laser into the cavity is amplified by rotational motion of the BEC, leading to so-called `rotational superradiance'---in a loose analogy to black-hole physics---with distinct critical scaling properties. In turn, cavity photons mediate long-range, periodic attractive interactions among the vortices, which compete with pair-wise logarithmic repulsive vortex interactions and deform the Abrikosov triangular vortex lattice favoring a stripe-like pattern. Remarkably, rotation of the BEC and topological properties of the vortex lattice can be monitored nondestructively through the cavity output field.

\end{abstract}

\maketitle

\section{Introduction}

Invariance under the time-reversal symmetry (TRS) indicates equivalence between two directions of motion in time~\cite{Gottfried2003}. For charged particles such as electrons, the TRS can be explicitly broken by, e.g., an applied external magnetic field. This leads to a wealth of intriguing phenomena, ranging from the integer and fractional quantum Hall effects to fractal energy spectra of Hofstadter's butterfly~\cite{QH1987}. Such phenomena are naturally absent in charge-neutral particles as they do not minimally couple to electromagnetic gauge potentials, and hence the TRS is respected. However, this restriction can be overcome and the TRS can also be broken explicitly in charge-neutral systems like ultracold atomic gases by making use of the Coriolis force in a rotating frame. Besides exploiting rotation, numerous other techniques exist now to mimic various types of minimal gauge couplings for neutral quantum particles and hence to break the TRS explicitly~\cite{Dalibard2011, Goldman2014}. 

Undoubtedly, the response of an atomic Bose--Einstein condensate (BEC) to a synthetic magnetic field piercing the BEC and breaking the TRS is intriguing: Quantized vortices form inside the BEC, where the phase of the condensate wave function has singularities in vortex cores and correspondingly the density falls to zero in these points~\cite{Fetter2001, Cooper2008, Fetter2009}. This is intimately related to the superfluid nature of the BEC and is reminiscent of the behavior of rotating superfluid helium. At low temperatures, these vortices are arranged in a triangular ``Abrikosov'' lattice~\cite{Tkachenko1966} due to pairwise logarithmic repulsive interactions between the vortices~\cite{Fetter1965Vortices_I, Castin1999}. Such vortices and vortex lattices have been created in BECs through different experimental techniques~\cite{Srinivasan2006}, including rotation and stirring~\cite{Matthews1999, Madison2000, AboShaeer2001,Haljan2001,Hodby2001,Engels2002, Schweikhard2004}, phase imprinting~\cite{Leanhardt2002,Andersen2006,Beattie2013}, and engineering a synthetic gauge field~\cite{Lin2009,Liu2024vortex}.

During the past two decades, there has been a tremendous effort to study the celebrated Dicke-superradiance and self-ordering phenomena in quantum gases inside optical cavities~\cite{Mivehvar2021}. Superradiance and corresponding self-organization in the original forms have already been experimentally observed with both bosonic~\cite{Baumann2010, Schmidt2014Dynamical, Klinder2015, Kollr2017, Zhiqiang2017} and fermionic~\cite{Zhang2021, Helson2023} quantum gases inside cavities. Many extensions to the original models have also been implemented experimentally~\cite{Lonard2017Supersolid, Guo2021An, Kongkhambut2022Observation}. On the theoretical side, numerous novel directions have been explored as well, including possibilities for simulating dynamical gauge potentials for quantum gases~\cite{Kollath2016, Zheng2016Superradiance, Ballantine2017Meissner-like, Colella2019, Colella2022}. Despite all these theoretical and experimental developments, one fundamental question has been completely overlooked: What happens to superradiance and self-ordering if the TRS is explicitly \emph{broken} (by a minimal-gauge-coupling-like effect)? In this article, we address this intriguing question in the context of many-body cavity QED and find a wealth of intriguing phenomena.

\begin{figure}[t!]
\centering
\includegraphics [width=0.35\textwidth]{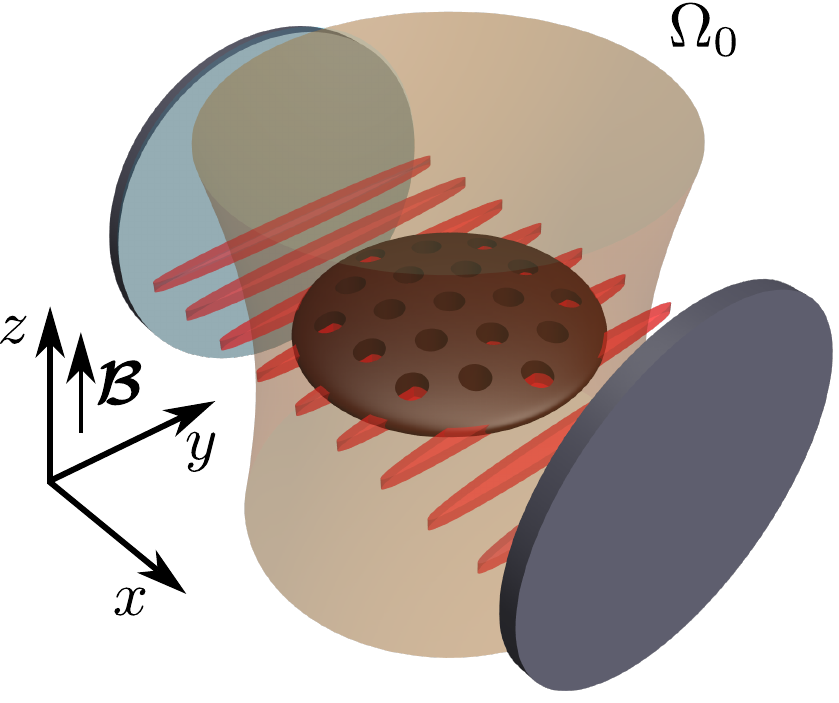}
\caption{Model: A harmonically-trapped 2D BEC (the dark brown disk), pierced with an external synthetic magnetic field $\boldsymbol{\mathcal{B}}=\mathcal{B}\mathbf{e}_z$ and coupled to a cavity mode (red stripes). The BEC is transversely driven along the $z$ axis by a pump laser (the light brown cylinder) with Rabi frequency $\Omega_0$.} 
\label{fig:scheme}
\end{figure}

We consider a two-dimensional (2D) BEC, harmonically trapped in the longitudinal plane of a standing-wave cavity, as shown in Fig.~\ref{fig:scheme}. The BEC is coupled to an off-resonant mode of the cavity and is further driven dispersively by a transverse pump laser oriented perpendicular to the plane of the BEC, thus ensuring the pump field has no position dependence in the BEC plane. A synthetic magnetic field pointing in the same direction as the pump laser pierces the BEC, hence breaking the TRS and altering the underlying nature of the system fundamentally. Our most notable finding is that the Dicke superradiant (SR) threshold decreases when the rotational energy of the BEC increases as a function of the increasing synthetic gauge field as shown in Fig.~\ref{fig:eta_c-B_c}(a). In other words, rotation of the BEC resulted from the broken TRS enhances SR scattering, reminiscent of the amplification of a wave scattered off a fast rotating absorbing body~\cite{ZelDovich1971Generation, ZelDovich1972Amplification}, such as rotating black holes~\cite{Misner1972Interpretation, PRESS1972Floating, Starobinskii1973, Konoplya2008}, at the expense of rotational energy~\cite{PENROSE1971}. Motivated by this analogy, we \emph{loosely} refer to the enhancement of Dicke SR scattering by rotation in our system as `\emph{rotational superradiance}'~\cite{Bekenstein1998The}. We also show that, in turn, the critical synthetic magnetic field for single-vortex stability in the BEC is decreased drastically in the SR regime as illustrated in Fig.~\ref{fig:eta_c-B_c}(b). This is due to the emergent cavity potential, which breaks the rotational symmetry and allows for the angular momentum to be pumped into the system via photon scattering processes. We then examine both the steady state and dynamical response of the system across these thresholds as shown in Figs.~\ref{fig:phase-diagrams} and~\ref{fig:dynamical_responses}. For chosen parameters, the SR phase transition as a function of the gauge field (i.e., rotation) exhibits a series of first-order transitions, while as a function of the pump has a second-order nature---but with a very different scaling behavior compared to the original SR phase transition. Another key finding is that cavity photons mediate long-range, periodic attractive interactions among \emph{vortices}, which compete with logarithmic repulsive vortex interactions and distort the Abrikosov triangular vortex lattice, favoring self-ordered striped vortex patterns.  

\section{Model}

Consider a harmonically-trapped 2D BEC in the $x$-$y$ plane, pierced with an \emph{external} uniform synthetic magnetic field $\boldsymbol{\mathcal{B}}=\mathcal{B}\mathbf{e}_z$ along the $z$ axis, as shown in Fig.~\ref{fig:scheme}. The BEC is located inside an optical cavity oriented along the $x$ axis and is coupled with strength $\mathcal{G}(x)=\mathcal{G}_0\cos(k_cx)$ to a single standing-wave mode of the cavity, with $k_c=2\pi/\lambda_c=\omega_c/c$ being the wavenumber of the mode. The BEC is further driven with rate $\Omega_0$ by a transverse pump laser with frequency $\omega_p$ along the $z$ axis. Both the cavity and pump fields are far off-resonance so that atomic excited states are not populated significantly and only low-lying momentum states are excited. In this regime the BEC order parameter $\psi(\mathbf{r},t;\boldsymbol{\mathcal{A}}\,,\alpha) = \sqrt{n}\exp(i \theta)$ obeys a generalized Gross--Pitaevskii equation (GPE)~\cite{Mivehvar2021},
\begin{align} \label{eq:BEC-order-parameter}
i\hbar\frac{\partial \psi}{\partial t}=\left[\frac{(\mathbf{p} - \boldsymbol{\mathcal{A}})^2}{2M}
+V(\mathbf{r},\alpha)+g_0 n \right]\psi,
\end{align}
where $M$ is the atomic mass, $g_0$ the strength of repulsive two-body contact interactions, and $\boldsymbol{\mathcal{A}}(\mathbf{r}) = \mathcal{B} (-y,x,0)/2$ the external time-independent synthetic vector potential in the symmetric gauge (with the synthetic charge set to one for simplicity) minimally coupled to the momentum operator $\mathbf{p}$, breaking the TRS of the system explicitly. This external gauge potential can readily be implemented via various available techniques~\cite{Goldman2014}, such as two-photon Raman processes~\cite{Lin2009} or rotation of the BEC~\cite{Matthews1999, Madison2000, AboShaeer2001,Haljan2001,Hodby2001,Engels2002, Schweikhard2004} which also results in an extra anti-trapping potential. For concreteness, we consider the former case here and present the latter case in detail in Appendix~\ref{app:rotating_BEC}.

The potential $V(\mathbf{r},\alpha)=V_{\rm tr}(\mathbf{r})+V_{\rm SR}(x,\alpha)$ is composed of two terms: A symmetric harmonic trap $V_{\rm tr}(\mathbf{r})=M\omega_{\rm tr}^2r^2/2$ with the trap frequency $\omega_{\rm tr}$ and an emerged, \emph{dynamical} cavity optical lattice $V_{\rm SR}(x,\alpha)=\hbar U |\alpha|^2 \cos^2(k_c x)+2\hbar\eta\text{Re}(\alpha)\cos(k_c x)$. Here, $\alpha(t;n)$ denotes the coherent amplitude of the cavity field determined self-consistently using the atomic density $n(\mathbf{r},t;\boldsymbol{\mathcal{A}}\,,\alpha)$ via~\cite{Mivehvar2021, Defenu2023Long},  
\begin{align} \label{eq:cavity-order-parameter}
i\frac{\partial \alpha}{\partial t}&=-\left(\Delta_c+i\kappa-U\int n(\mathbf{r})\cos^2(k_cx)\, dxdy \right)\alpha
\nonumber\\
&+\eta\int n(\mathbf{r})\cos(k_cx)\, dxdy,
\end{align}
where $\Delta_c=\omega_p-\omega_c$ is the pump-cavity detuning, $\kappa$ the field decay rate, $U=\mathcal{G}_0^2/\Delta_a$, and $\eta=\mathcal{G}_0\Omega_0/\Delta_a$ (with $\Delta_a$ being the detuning of the pump laser from an atomic excited state). Since the dynamics along $z$ is completely frozen due to the strong confinement, the potential arising from the pump laser in the $x$-$y$ plane is merely a constant, $\Omega_0^2/\Delta_a$, and has thus been omitted in $V(\mathbf{r},\alpha)$. 

In the following, we study the intricate interplay of superradiance and the synthetic background gauge potential (or equivalently ``rotation'') in our system and show that a wealth of intriguing phenomena appears. This is due to the fact that, on the one hand the atomic density $n(\boldsymbol{\mathcal{A}}\,,\alpha)$ depends directly on the gauge potential $\boldsymbol{\mathcal{A}}$ and the cavity field $\alpha$, while on the other hand the cavity field $\alpha(n)$ depends directly on the atomic density $n$ and hence indirectly on the gauge potential $\boldsymbol{\mathcal{A}}$. First, we focus on the steady-state physics of the system, by setting $\partial_t\psi=-i\mu\psi/\hbar$ (with $\mu$ being the chemical potential) and $\partial_t\alpha=0$ in Eqs.~\eqref{eq:BEC-order-parameter} and~\eqref{eq:cavity-order-parameter}, respectively. Afterwards, nonequilibrium dynamics of the system are examined through the full coupled equations of motion, Eqs.~\eqref{eq:BEC-order-parameter} and~\eqref{eq:cavity-order-parameter}. We refer interested readers to Appendix~\ref{app:SM-numerics-details} for the details of numerical calculations. 

Throughout this paper, the units for length, frequency, and energy are given by $a_{\rm ho} = \sqrt{\hbar/(M\omega_{\rm tr})}$, $\omega_{\rm tr}$ and $\hbar\omega_{\rm tr}$ respectively. We also introduce the parameter $\zeta = a_{\rm ho}/\lambda_c$ characterizing the ratio between the size of the BEC and the wavelength of the cavity mode.

\section{Results and Discussion}

\subsection{Criticalities}

\begin{figure}[t!]
\centering
\includegraphics [width=0.46\textwidth]{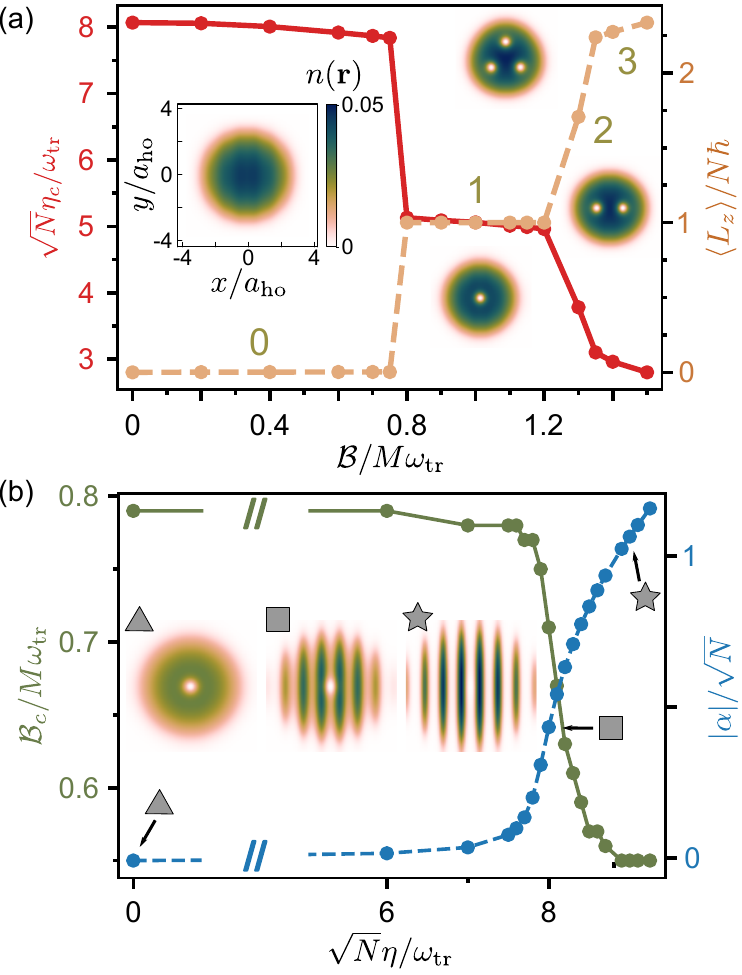}
\caption{Critical behavior of the system: The SR threshold $\sqrt{N}\eta_c$ (the red line with respect to the left axis) and the average of the $z$ component of the angular momentum per particle $\langle L_z\rangle/N\hbar$ (the orange dashed line with respect to the right axis) as a function of the synthetic magnetic field $\mathcal{B}$~(a), and the critical gauge field $\mathcal{B}_c$ (the olive-green curve with respect to the left axis) and the absolute value of the rescaled cavity-field amplitude $|\alpha|/\sqrt{N}$ (the blue dashed curve with respect to the right axis) as a function of the pump strength $\sqrt{N}\eta$~(b).  Numbers in~(b) indicate the amount of vortices in the steady-state BEC density profile $n(\mathbf{r})$ and the insets show typical densities $n(\mathbf{r})$ for indicated parameters. The other parameters are set to $(NU,\Delta_c,\kappa) = (-1,-5,2)\,\omega_{\rm tr}$ and $N g_0 = 150\hbar\omega_{\rm tr} a_{\rm ho}^2/\zeta^2$ with $\zeta = 0.9$.} 
\label{fig:eta_c-B_c}
\end{figure}

Let us start by examining the critical behavior of the system in the vicinity of the SR phase transition and the single-vortex nucleation. We illustrate the SR threshold $\eta_c(\mathcal{B})$ as a function of the synthetic magnetic field in Fig.~\ref{fig:eta_c-B_c}(a), and the critical synthetic magnetic field $\mathcal{B}_c(\eta)$ for the single vortex nucleation as a function of the pump laser in Fig.~\ref{fig:eta_c-B_c}(b); see Appendix~\ref{app:sr_threshold} for numerical details.

The SR threshold $\eta_c(\mathcal{B})$ shows an intriguing behavior: It decreases as the gauge field $\mathcal{B}$ increases, exhibiting discrete drops in specific values of $\mathcal{B}$. These plunges correspond to changes in the number of vortices in the system, quantified by the average of the $z$ component of the angular momentum operator $L_z=xp_y-yp_x$, as also shown in Fig.~\ref{fig:eta_c-B_c}(a) with respect to the right axis. Such behavior indicates that rotation of the BEC driven by $\mathcal{B}$ enhances the tendency of SR photon scattering~\cite{Lochan2020Detecting}. This is reminiscent of the amplification of a wave scattered off a fast rotating absorbing body at the expense of rotational energy, first predicted by Zel'dovich for a rotating conducting cylinder~\cite{ZelDovich1971Generation, ZelDovich1972Amplification} and then extended to rotating black holes~\cite{Misner1972Interpretation, PRESS1972Floating, Starobinskii1973, Konoplya2008}---all in a close analogy to the Penrose mechanism for extracting rotational energy from a rotating black hole by a \emph{massive body}~\cite{PENROSE1971}. A similar effect has also been predicted for sound waves in flowing (quantum) fluids containing vortices~\cite{Basak2003, Berti2004Quasinormal, Federici2006Superradiance, Ghazanfari2014Acoustic, Richartz2015Rotating, Cardoso2016Detecting, Marino2020Zero, Giacomelli2021Understanding, Cardoso2022}, where a vortex can be imagined as a sonic black hole analog~\cite{Unruh1981Experimental, Unruh1995Sonic, Lahav2010Realization}. Indeed, such rotational superradiance of sound waves has been recently observed in water with a draining vortex~\cite{Torres2017Rotational} and in air with a rotating absorbing disk~\cite{Cromb2020}. In a close analogy to these phenomena, we \emph{loosely} designate the enhanced SR photon scattering due to the gauge field $\mathcal{B}$ as `\emph{rotational superradiance}'~\cite{Bekenstein1998The}. 

This amplification in our system can be attributed on the one hand to density inhomogeneities caused by vortices, which only \emph{minimally} seed the SR photon scattering as shown in Appendix~\ref{app:vortex_inhom}. On the other hand and most importantly, the density response of the system in the presence of $\mathcal{B}$ is determined by Landau-like states~\cite{Appugliese2022Breakdown}, which is fundamentally different from the response of a BEC to usual self-ordering in plane-wave or Bloch basis~\cite{Zupancic2019P-Band}; see Appendix~\ref{app:sr_threshold} for more details.

The critical magnetic field $\mathcal{B}_c(\eta)$ for a vortex to be energetically favored and nucleate inside the BEC also exhibits interesting features, as depicted in Fig.~\ref{fig:eta_c-B_c}(b). Although $\mathcal{B}_c(\eta)$ is almost independent of the pump power in the normal phase $\sqrt{N}\eta\lesssim6\omega_{\rm tr}$, it reduces significantly by increasing $\eta$ in the SR state. This is due to the emerged intracavity potential $V_{\rm SR}(x,\alpha)$ which creates density modulation for the BEC, thus decreasing the energy barrier for a vortex to enter the BEC through a density minimum similar to a recent experiment with a dipolar supersolid~\cite{Casotti2024Observation}. In the deep SR phase, the BEC separates into disjoint 1D tubes and hence $\mathcal{B}_c$ almost saturates.


\begin{figure}[t!]
\centering
\includegraphics [width=0.48\textwidth]{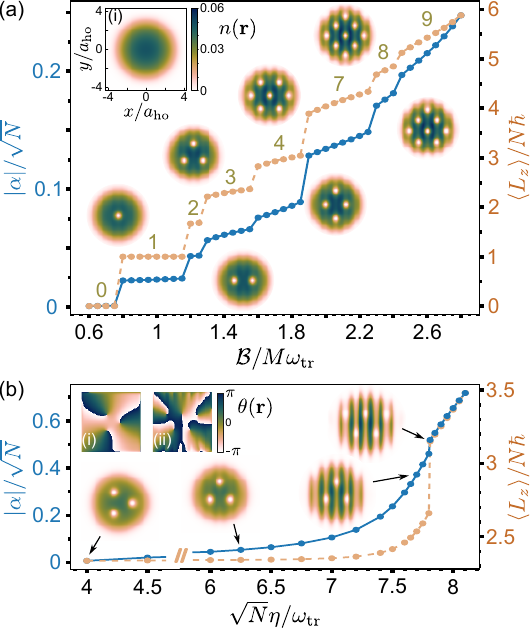}
\caption{SR phase transition: 
The steady-state field amplitude $|\alpha|/\sqrt{N}$ (blue line with respect to the left axis) and the angular momentum $\langle L_z\rangle/N\hbar$ (orange dashed line with respect to the right axis) as functions of the synthetic gauge field $\mathcal{B}$ for the fixed pump strength $\sqrt{N}\eta = 6.5\omega_{\rm tr}$~(a), and of $\sqrt{N}\eta$ for fixed $\mathcal{B} = 1.5M\omega_{\rm tr}$~(b). The inset~(a.i) shows the BEC density profile $n(\mathbf{r})$ for~$\mathcal{B} = 0.6M\omega_{\rm tr}$, while~(b.i-ii) illustrate the BEC phase profile $\theta(\mathbf{r})$ for $\sqrt{N}\eta = 7\omega_{\rm tr}$ and $7.81\omega_{\rm tr}$, respectively. All the other insets in~(a) and~(b) depict typical $n(\mathbf{r})$ for indicated parameters. The other parameters are the same as Fig.~\ref{fig:eta_c-B_c}.} 
\label{fig:phase-diagrams}
\end{figure}

\subsection{Steady-state phase diagrams}

We now examine the steady-state behavior of the system across the SR phase transition. Let us commence by looking at the response of the system at a fixed $\eta$ as a function of increasing $\mathcal{B}$. We fix the value of the pump strength at $\sqrt{N}\eta = 6.5\omega_{\rm tr}$, which is below the SR threshold in the absence of the gauge field $\sqrt{N}\eta_c(\mathcal{B}=0)\approx 8.1\omega_{\rm tr}$; see Fig.~\ref{fig:eta_c-B_c}(a). As can be seen from Fig.~\ref{fig:phase-diagrams}(a), for $\mathcal{B} < 0.8M\omega_{\rm tr}$ the BEC density $n(\mathbf{r})$ contains no vortex [i.e., $\langle L_z \rangle=0$ as indicated by the right axis of Fig.~\ref{fig:phase-diagrams}(a)] and there is no real photon inside the cavity. The BEC density follows closely the Thomas-Fermi distribution as depicted in Fig.~\ref{fig:phase-diagrams}(a.i) for~$\mathcal{B} = 0.6M\omega_{\rm tr}$.

When $\mathcal{B} \gtrsim 0.8M\omega_{\rm tr}$, it becomes energetically favorable for a vortex to nucleate in the center of the BEC and correspondingly the average angular momentum $\langle L_z \rangle$ jumps from zero to one. Concurrently, the BEC scatters constructively some small amount of photons from the pump into the cavity, which in turn modulates slightly the BEC density due to the emergent $V_{\rm SR}(x,\alpha)$; see the density inset above the $\langle L_z \rangle=1$ plateau in Fig.~\ref{fig:phase-diagrams}(a). This is consistent with the SR-threshold $\eta_c(\mathcal{B})$ behavior shown in Fig.~\ref{fig:eta_c-B_c}(a): By increasing $\mathcal{B}$ beyond $0.8M\omega_{\rm tr}$, the SR threshold reduces drastically from approximately $7.8\omega_{\rm tr}$ to $5.1\omega_{\rm tr}$. The chosen pump strength $\sqrt{N}\eta = 6.5\omega_{\rm tr}$ then lies above the threshold and the system, therefore, enters the SR phase. Note that the SR order parameter $\alpha$ exhibits a discrete jump across the phase transition, revealing its first-order nature.          

By further increasing $\mathcal{B}$ more vortices appear in the BEC, as shown by the density insets and quantified by increasing $\langle L_z \rangle$. As expected, $\langle L_z \rangle$ is no longer quantized at integer values~\cite{Butts1999Predicted}. Although the pump strength is fixed $\sqrt{N}\eta = 6.5\omega_{\rm tr}$, the BEC scatters more photons into the cavity due to increased rotational motion of the cloud and density inhomogeneities, hence further justifying the chosen name `rotational superradiance'. Equivalently, this can be explained in another complementary way: By increasing $\mathcal{B}$, the SR threshold $\eta_c(\mathcal{B})$ decreases further [see Fig.~\ref{fig:eta_c-B_c}(a)] and, therefore, for a fixed pump strength $\eta$ the system goes deeper into the SR regime. It is interesting to note that the cavity field~$\alpha$ exhibits a series of step-like first-order transitions as $\mathcal{B}$ increases, in complete accordance with $\langle L_z \rangle$.

We now inspect the behavior of the system at a fixed $\mathcal{B}$ as a function of the pump strength $\eta$. The gauge field is fixed at $\mathcal{B}=1.5M\omega_{\rm tr}$, which stabilizes three vortices inside the BEC density in the absence of the cavity field, as shown by the leftmost density inset of Fig.~\ref{fig:phase-diagrams}(b). For weak pump strengths $\sqrt{N}\eta\lesssim4.5\omega_{\rm tr}$, the system is in the normal state as shown in Fig.~\ref{fig:phase-diagrams}(b). By increasing $\eta$, the SR phase sets in. Unlike the SR phase transition resulted from increasing $\mathcal{B}$, the SR phase transition here as a function of the increasing pump strength $\eta$ is second order. However, the scaling behavior $|\alpha(\mathcal{B}=1.5M\omega_{\rm tr})|\propto(\eta-\eta_c)$ obtained numerically by fitting the cavity field above the SR threshold in our finite-size system, see Appendix~\ref{app:scailing}, is different from the common SR scaling $|\alpha(\mathcal{B}=0)|\propto(\eta-\eta_c)^{1/2}$~\cite{Nagy2011Critical,Kirton2018}. This implies that $\mathcal{B}$ changes the universality class of the SR phase transition, at least in finite-size systems.

Interestingly, the average angular momentum $\langle L_z \rangle$ also grows continuously in the SR regime with increasing $\eta$, since the emerged cavity potential $V_{\rm SR}(x,\alpha)$ breaks the rotational symmetry of the system, allowing for the angular momentum to be pumped into the system via photon scattering processes. Although $\mathcal{B}$ is fixed, deep in the SR regime two more vortices remarkably nucleate in the BEC density [see the rightmost density inset of Fig.~\ref{fig:phase-diagrams}(b)], indicated by the discrete jump in $\langle L_z \rangle$ around $\sqrt{N}\eta\approx7.8\omega_{\rm tr}$. Likewise, this can be accounted for by noting Fig.~\ref{fig:eta_c-B_c}(b): By increasing the pump strength, $\mathcal{B}_c(\eta)$ decreases substantially in the SR regime and, therefore, for a fixed $\mathcal{B}$ more vortices are favored in the BEC density. Note also that the discrete jump in $\langle L_z \rangle$ is accompanied by a small discrete jump in $\alpha$.

\subsection{Vortex properties and nondestructive probing}

In the absence of the cavity field, vortices in the BEC form an Abrikosov triangular lattice owing to pairwise logarithmic repulsive interactions $\propto\sum_{j>i}\ln{|\mathbf{r}_i-\mathbf{r}_j|}$ between the vortices~\cite{Cooper2008}. Similar to cavity-mediated long-range inter-atomic interactions, cavity photons also mediate long-range periodic attractive interactions $\propto\eta^2\sum_{i,j}\cos(k_cx_i)\cos(k_cx_j)$ among the \emph{vortices} as well as repulsive interactions between \emph{atoms} and \emph{vortices}, as we show below. These cavity-mediated interactions can be obtained by integrating out the photonic degree of freedom and rewriting the Hamiltonian only in terms of atomic operators. The cavity-induced effective interaction Hamiltonian then takes the following form,
\begin{equation}
\label{eq:at_at_int}
    H_{\rm eff-int} = \iint\mathcal{D}(x,x')n(\mathbf{r})n(\mathbf{r'})d\mathbf{r}d\mathbf{r'},
\end{equation}
where
\begin{equation}
\mathcal{D}(x,x') = \frac{2\hbar\Delta_c \eta^2} {\Delta_c^2 + \kappa^2}\cos(k_c x)\cos(k_c x'). 
\end{equation} 

By decomposing the total atomic density $n(\mathbf{r}) = n_0(\mathbf{r}) - n_v(\mathbf{r})$ into the vortex-free part $n_0(\mathbf{r})$ and the vortex density $n_v(\mathbf{r})$ (note the minus sign preceding $n_v$ which accounts for the `holes' in the total density $n$), the cavity-induced interaction Hamiltonian~\eqref{eq:at_at_int} can be recast as,
\begin{align}
\label{eq:vor_at_int}
    H_{\rm eff-int} &= \iint\mathcal{D}(x,x')n_0(\mathbf{r})n_0(\mathbf{r'})d\mathbf{r}d\mathbf{r'} \nn\\
    &+  \iint\mathcal{D}(x,x')n_v(\mathbf{r})n_v(\mathbf{r'})d\mathbf{r}d\mathbf{r'} \nn\\
    &- 2\iint\mathcal{D}(x,x')n_0(\mathbf{r})n_v(\mathbf{r'})d\mathbf{r}d\mathbf{r'}.
\end{align}
Here, the first term corresponds to the conventional cavity-mediated, long-range periodic inter-atomic interactions. The second term represents all-to-all periodic interactions among vortices, mediated by cavity photons. The last term in Eq.~\eqref{eq:vor_at_int} can be interpreted as an interaction between atoms and vortices---or matter and holes. The character of these interactions is determined by the sign of the cavity detuning $\Delta_c$.  For our chosen parameters where $\Delta_c<0$, while the first two interactions are attractive, the last interaction between atoms and vortices is repulsive.

The interplay between logarithmic repulsive and cavity-mediated attractive inter-vortex interactions, the contact collisional and cavity-mediated inter-atomic interactions, as well as cavity-mediated atom-vortex interactions determines single vortex properties as well as the vortex-lattice structure. For a rotating BEC, the vortex core size is intimately related to the healing length~\cite{Fischer2003,Baym2004}. In the Thomas-Fermi limit the vortex core radius grows by increasing $\Omega_{\rm rot}$, as rotation reduces the condensate density due to the anti-trapping potential $W_{\rm rot}(\mathbf{r})=-M\Omega_{\rm rot}^2(x^2 + y^2)/2$. In the case of an external synthetic magnetic field $\mathcal{B}$, the anti-trapping potential is, however, absent, and the vortex core size is less affected by $\mathcal{B}$. Therefore, in our system, any change in the vortex core structure can be attributed to the cavity-mediated interactions, Eq.~\eqref{eq:vor_at_int}.

\begin{figure}[t!]
\centering
\includegraphics[width=0.48\textwidth]{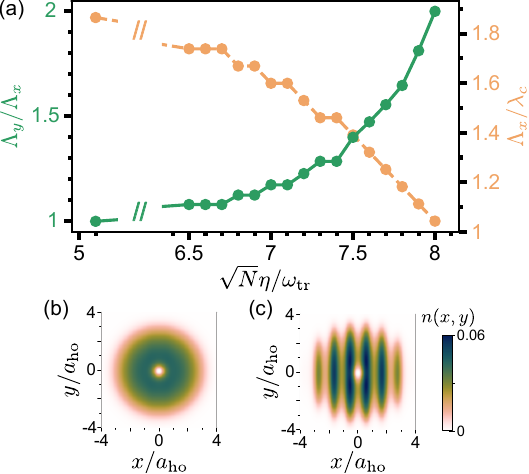}
\caption{Vortex core structure: The ratio between the vortex size along the $y$ axis to the vortex size along the $x$ axis, $\Lambda_y/\Lambda_x$, (green solid curve with respect to the left axis) and the vortex size along the $x$ axis to the wavelength of the cavity mode, $\Lambda_x/\lambda_c$, (orange dashed line with respect to the right axis) as functions of the pump strength $\sqrt{N}\eta$ for the fixed synthetic magnetic field $\mathcal{B} =  M\omega_{\rm tr}$ (a). By increasing the pump strength, the cavity-mediated interactions become stronger and lead to a notable squeezing of the vortex core along the $x$ direction. The density profiles for $\sqrt{N}\eta = 5.1\omega_{\rm tr}$ (b) and $\sqrt{N}\eta = 8\omega_{\rm tr}$ (c). Note how the vortex is deformed in panel (c) by the cavity-mediated interactions. The other parameters are the same as in Fig.~\ref{fig:eta_c-B_c}.}
\label{fig:SM_vortex_size}
\end{figure}

In Fig.~\ref{fig:SM_vortex_size}, we show the ratio between the vortex size along the $y$ axis to the vortex size along the $x$ axis, $\Lambda_y/\Lambda_x$, for the increasing value of the pump strength $\sqrt{N}\eta$ at the fixed value of the synthetic magnetic field $\mathcal{B} = M\omega_{\rm tr}$, corresponding to a single vortex at the center of the BEC. As a vortex size along a given axis, we take the distance between the two density maxima on the opposite sides of the vortex. For small pump strengths, the cavity-mediated interactions~\eqref{eq:vor_at_int} are small and, hence, the vortex is almost symmetric $\Lambda_y/\Lambda_x \approx 1$. By increasing the pump strength, the cavity-mediated interactions become stronger. In particular, the cavity-mediated attractive inter-atomic interactions favor a $\lambda_c$-periodic density modulation along the $x$ direction, where $\lambda_c$ is the cavity wavelength. On the other hand, since the cavity-mediated atom-vortex interactions are repulsive, they lead to $\lambda_c$/2 separation between the atomic density and the vortex (in general, vortices) along the $x$ direction. Therefore, the vortex tends to localize within a density minimum, in order to minimize the cavity-mediated atom-vortex interaction energy. In the deep SR regime, the vortex is strictly localized within a density minimum, resulting in a significant squeezing of the vortex along the $x$ axis as can be seen from Fig.~\ref{fig:SM_vortex_size}. The compression of the vortex along the $x$ direction within a density minimum can readily be seen by the ratio $\Lambda_x/\lambda_c$, as also shown in Fig.~\ref{fig:SM_vortex_size} with respect to the right axis.

The vortex-lattice structure is also modified under the influence of the cavity-mediated interactions, Eq.~\eqref{eq:vor_at_int}. When the cavity-mediated periodic inter-vortex interactions become strong enough to overcome the logarithmic repulsive vortex interactions, the triangular vortex lattice deforms into a stripe-like pattern as can be seen from the density insets of Figs.~\ref{fig:phase-diagrams}(a) and (b), reminiscent of vortex-lattice deformation in dipolar quantum gases~\cite{Roccuzzo2020, Klaus2022}.

Let us reiterate that the behavior of the cavity field as a function of the gauge field and the pump laser, $\alpha(\mathcal{B})$ and $\alpha(\eta)$, bears a strong resemblance to that of the average angular momentum $\langle L_z \rangle$ of the BEC. In particular, discrete jumps in $\langle L_z\rangle$, which indicate \emph{topological} deformation of the vortex lattice, are accompanied by discrete jumps in the cavity field. Therefore, this provides the possibility for the nondestructive monitoring of topological properties of a vortex lattice in a BEC and quantum-Hall-like states through the cavity output field~\cite{Brennecke2013, Kumar2021, Skulte2023Quantum}. 

\subsection{Nonequilibrium dynamics}

\begin{figure}[t!]
\centering
\includegraphics [width=0.48\textwidth]{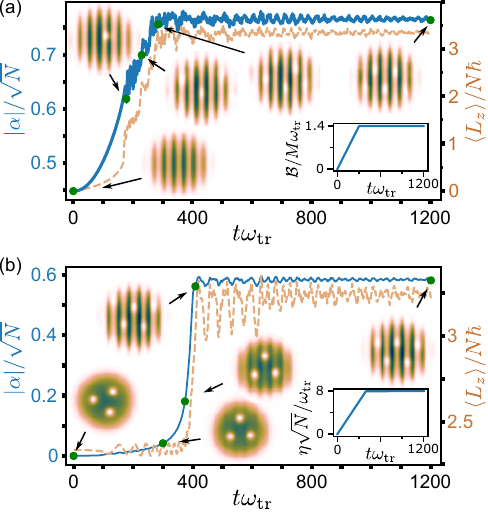}
\caption{Nonequilibrium dynamics of the system: The time evolution of $|\alpha|/\sqrt{N}$ (blue curves with respect to the left axis) and $\langle L_z\rangle/N\hbar$ (orange dashed curves with respect to the right axis) after the linear ramp of $\mathcal{B}$ from 0 to $1.4M\omega_{\rm tr}$ during initial $300/\omega_{\rm tr}$ seconds (see the inset in panel a) at fixed $\sqrt{N}\eta = 8.2\omega_{\rm tr}$~(a), and $\sqrt{N}\eta$ from 0 to $7.9\omega_{\rm tr}$ during initial $400/\omega_{\rm tr}$ seconds (see the inset in panel b) at fixed $\mathcal{B} = 1.5M\omega_{\rm tr}$~(b). Typical BEC density profiles $n(\mathbf{r})$ are shown for indicated times. The other parameters are the same as Fig.~\ref{fig:eta_c-B_c}.}
\label{fig:dynamical_responses}
\end{figure}

Finally, we examine the nonequilibrium dynamics of the system following the ramp of the system parameters $\eta$ and $\mathcal{B}$, respectively. To this end, we start with the steady state of the system for fixed $\sqrt{N}\eta = 8.2\omega_{\rm tr}$ and no gauge field $\mathcal{B} = 0$ [cf.\ Fig.~\ref{fig:eta_c-B_c}(a)]. The steady state is a density-modulated striped SR state as shown by the first density inset in Fig.~\ref{fig:dynamical_responses}(a). The gauge field $\mathcal{B}$ is then ramped linearly from zero to $1.4M\omega_{\rm tr}$ during the first $300/\omega_{\rm tr}$ seconds and kept constant afterward, see the inset in Fig.~\ref{fig:dynamical_responses}(a) depicting this protocol. As expected, $\langle L_z \rangle$ increases and accordingly vortices enter the BEC, especially through density minima as shown by the density insets; follow the link in Ref.~\cite{anim2024} for an animation of the real-time dynamics. Remarkably, the system quickly cools down to almost its steady state via the cavity dissipation channel. Small amplitude, high-frequency oscillations stem from atomic excitations that do not couple to the cavity field, such as kinetic-energy excitations in the $y$ direction. The dynamical response of the system also reveals the rotational superradiance where photon scattering into the cavity field $\alpha$ is amplified by almost 70\% owing to the rotational motion of the BEC. Note that, however, the density distortion due to the nucleated vortices is just minimal as the vortices sit in the already existing BEC density minima. This hints that this SR amplification should originate chiefly from the fundamental change of the response function of the system due to the TRS-breaking gauge field $\mathcal{B}$. As Fig.~\ref{fig:dynamical_responses}(a) illustrates, the nonequilibrium dynamics of the BEC during the vortex nucleation can be monitored nondestructively and faithfully via the cavity-output field.  

We carry out similar analyses for the linear ramp of the pump strength at fixed $\mathcal{B}$ as depicted in Fig.~\ref{fig:dynamical_responses}(b). The BEC is initially in the normal phase containing three vortices as shown by the first density inset in Fig.~\ref{fig:dynamical_responses}(b). Consequently, the system evolves into the SR phase characterized by nonzero $\alpha$ and emergent density modulation. It is interesting to note that in the early stage of the dynamics, the three vortices rearrange clearly due to the emergent cavity-mediated inter-vortex interactions. As expected, angular momentum is also pumped into the system, and two additional vortices nucleate; see Ref.~\cite{anim2024} for an animation of the dynamics.


\section{Conclusion and Outlook}

We showed that breaking the TRS has fundamental consequences in the SR phase transition and self-ordering. In particular, steady-state superradiance can be amplified by rotation in our setup, although the incident pump field has no angular momentum as in the original model~\cite{ZelDovich1971Generation, ZelDovich1972Amplification}. 
Our proposal can readily be implemented in any quantum-gas--cavity-QED laboratory~\cite{Baumann2010, Schmidt2014Dynamical, Klinder2015, Kollr2017, Zhiqiang2017,Zhang2021, Helson2023} by incorporating well-established techniques for breaking the TRS and creating vortices in quantum gases~\cite{Matthews1999, Madison2000, AboShaeer2001,Haljan2001,Hodby2001,Engels2002, Schweikhard2004,Leanhardt2002,Andersen2006,Beattie2013,Lin2009}.
Many intriguing fundamental questions still remain to be explored in our model, including hysteresis effects, the nature of polaritons and collective excitations, the accuracy of rotation measurements, and a pump field with orbital angular momentum~\cite{Baio2020, Baio2022} where it has already been shown to lead to \emph{transient} superradiance in free space~\cite{You2011,Robb2012,Das2016,Bachelard2022}, though the dependence of this transient superradiance and its threshold and criticality on the rotational motion of the BEC has not been explored. It is also interesting to extend this model to multi-mode cavity settings, where cavity-mediated interactions can be shaped due to multi-mode superradiance~\cite{Vaidya2018Tunable, Mivehvar2019Emergent, Masalaeva2023}. We leave addressing these interesting questions to future works.


\begin{acknowledgments}

We acknowledge inspiring discussions with Karol Gietka, Helmut Ritsch, Thomas Bland, Russell Bisset, Tobias Donner, and Thorsten Ackemann. This research was funded in whole or in part by the Austrian Science Fund (FWF) [grant DOIs: 10.55776/P35891 and 10.55776/P36850]. For open access purposes, the authors have applied a CC BY public copyright license to any author accepted manuscript version arising from this submission. F.\,M.~is also supported financially during a part of this work by the Tyrolean Science Promotion Fund (TWF) and the ESQ-Discovery Grant of the Austrian Academy of Sciences (\"OAW). 

\end{acknowledgments}

\appendix

\section{Steady-state phase diagram of the rotating BEC-cavity system}
\label{app:rotating_BEC}

As was mentioned in the main text, the external gauge potential $\boldsymbol{\mathcal{A}}$ can also be implemented by rotating the whole system (i.e., BEC plus the cavity) with angular velocity $\boldsymbol{\Omega}_{\rm rot}=\Omega_{\rm rot}\mathbf{e}_z$ around the $z$ axis. In the rotating frame, the 2D BEC order parameter $\psi(\mathbf{r},t;\Omega_{\rm rot},\alpha) = \sqrt{n}\exp(i \theta)$ is then determined via the generalized Gross--Pitaevskii equation,
\begin{equation}
\label{app_GPE}
i\hbar\frac{\partial \psi}{\partial t}=
\left[\frac{\mathbf{p}^2}{2M} +V(\mathbf{r},\alpha) -\Omega_{\rm rot} L_z + g_0 n\right]\psi,
\end{equation} 
where $L_z=(xp_y-yp_x)$ is the $z$ component of the angular momentum operator. The potential $V(\mathbf{r},\alpha)=V_{\rm tr}(\mathbf{r})+V_{\rm SR}(x,\alpha)$ now consists of the harmonic trap potential $V_{\rm tr}(x,y) =M\omega_{\rm tr}^2(x^2+\epsilon y^2)/2$ with the trap frequency $\omega_{\rm tr}$ and a weak anisotropy $\epsilon$, and an emerged, dynamical cavity optical lattice $V_{\rm SR}(x,\alpha)=\hbar U |\alpha|^2 \cos^2(k_c x)+2\hbar\eta\text{Re}(\alpha)\cos(k_c x)$ as before. 

The Gross--Pitaevskii equation~\eqref{app_GPE} can be recast as,
\begin{equation} 
\label{app_Ham_with_A_W}
i\hbar\frac{\partial \psi}{\partial t}=
\left[\frac{(\mathbf{p} - \boldsymbol{\mathcal{A}})^2}{2M} + V(\mathbf{r},\alpha)+ W_{\rm rot}(\mathbf{r})+g_0 n\right]\psi,
\end{equation}
where $\boldsymbol{\mathcal{A}}(\mathbf{r}) = M\Omega_{\rm rot}(-y,x,0)$ is an effective vector gauge potential and $W_{\rm rot}(\mathbf{r}) = -M\Omega_{\rm rot}^2(x^2 + y^2)/2$ an anti-trapping (centrifugal) potential. As one can see, this additional anti-trapping potential $W_{\rm rot}(\mathbf{r})$ is the only difference between the system rotating around the $z$ axis and the BEC pierced with an effective magnetic field $\boldsymbol{\mathcal{B}}=\mathcal{B}\mathbf{e}_z$, described by Eqs.~\eqref{app_ss}. In the following, we investigate whether this additional anti-trapping potential $W_{\rm rot}(\mathbf{r})$ has any major effect in the SR phase transition. 

\begin{figure}[t!]
\centering
\includegraphics[width=0.48\textwidth]{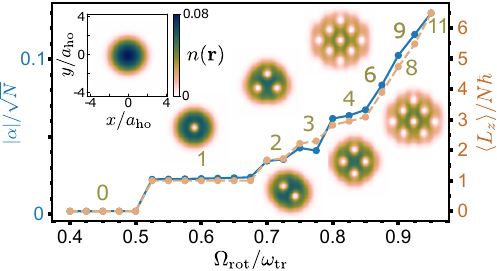}
\caption{Rotational SR in the rotating BEC-cavity system: The steady-state field amplitude $|\alpha|/\sqrt{N}$ (blue line with respect to the left axis) and the $z$ component of the average angular momentum per particle $\langle L_z\rangle/N\hbar$ (orange dashed line with respect to the right axis) as a function of the angular velocity $\Omega_{\rm rot}$ for the fixed pump strength $\sqrt{N}\eta = 4.5\omega_{\rm tr}$. The inset in the left top corner shows the BEC density profile $n(\mathbf{r})$ for~$\Omega_{\rm rot} = 0.4\omega_{\rm tr}$. All the other insets depict typical $n(\mathbf{r})$ for the indicated number of vortices in the obtained steady states. The other parameters are set to $(NU,\Delta_c,\kappa) = (-0.8,-5,2)\,\omega_{\rm tr}$, $N g_0 = 50\hbar\omega_{\rm tr} a_{\rm ho}^2/\zeta^2$ with $\zeta = 0.75$, and $\epsilon = 1.03$.}
\label{fig:SM_scan_along_omega}
\end{figure}

To obtain the steady-state phase diagram, similar to the one presented in Fig.~\ref{fig:phase-diagrams}(a), we perform a self-consistent imaginary-time evolution of the system as explained in the previous section. The results are presented in Fig.~\ref{fig:SM_scan_along_omega} where we have fixed the value of the transverse pump strength $\sqrt{N}\eta = 4.5\,\omega_{\rm tr}$ below the SR threshold in the absence of rotation, such that for small values of $\Omega_{\rm rot}$ the BEC is in the normal phase. As can be seen from Fig.~\ref{fig:SM_scan_along_omega}, for $\Omega_{\rm rot} \leqslant 0.5\omega_{\rm tr}$ the BEC density $n(\mathbf{r})$ follows closely the Thomas-Fermi distribution containing no vortices and the cavity field is in the vacuum state.

However, for slightly faster rotation $\Omega_{\rm rot} \gtrsim 0.5\omega_{\rm tr}$, the solution containing a vortex becomes stable, and the cavity field $\alpha$ is correspondingly populated. By further increasing the angular velocity $\Omega_{\rm rot}$ more vortices appear in the BEC and, consequently, more photons are scattered into the cavity mode. As expected, no fundamental change occurs due to the anti-trapping potential $W_{\rm rot}(\mathbf{r})$, except that the BEC expands as it rotates faster. Therefore, our proposed model for the rotational superradiance can be implemented in different setups. 

\section{Details of numerical calculations}
\label{app:SM-numerics-details}

To find the steady state of the system we self-consistently solve the following system of coupled equations 
\begin{align} \label{app_ss}
    i\hbar\frac{\partial \psi}{\partial t}&=
    \left[\frac{(\mathbf{p} - \boldsymbol{\mathcal{A}})^2}{2M}
+V(\mathbf{r},\alpha)+g_0 n \right]\psi =\mu \psi,\nonumber\\
    \alpha_{\rm ss}&=\frac{\eta\int n(\mathbf{r})\cos(k_cx)\, dxdy}{\Delta_c+i\kappa-U\int n(\mathbf{r})\cos^2(k_cx)\, dxdy},
\end{align}
using imaginary-time evolution. Taking into account the highly non-convex energy landscape of the system with many local minima~\cite{Isoshima1999,Castin1999,Feder1999}, we perform simulations with various initial conditions for each given set of parameters, and then select the solution with the lowest total energy including the free cavity-photon energy $-\Delta_c|\alpha|^2$. We find that in the SR regime the lowest energy solutions always correspond to the case where the SR potential $V_{\rm SR}(x,\alpha)$ is (positive) maximum in the center $x=0$. Stationary solutions are verified subsequently via real-time evolution, by adding some initial perturbations to the solutions.

As initial states, we take different modifications of the Thomas-Fermi wave function $\psi_{\rm TF} (\mathbf r) = \sqrt{[\mu_{\rm TF} -V_{\rm tr}(\mathbf r)]/g_0}$ with $\mu_{\rm TF}$ being the 2D Thomas-Fermi chemical potential~\cite{Ruan2016}:
\begin{align}
    \psi_{\rm rp} (\mathbf{r}) &= \psi_{\rm TF} (\mathbf r)e^{2\pi i \mathcal{R}(\mathbf{r})},\nn\\
    \psi_{\rm sv} (\mathbf{r}) &= \psi_{\rm TF} (\mathbf r)e^{i q \arctan(y/x)},\nn\\
    \psi_{\rm mv} (\mathbf{r})&= \psi_{\rm TF} (\mathbf r)\sum_{q} c_q e^{i q \arctan(y/x)}, 
\end{align}
where $\mathcal{R}(\mathbf{r})$ is a random number at position $\mathbf{r}$ and $q$ is the charge of the vortex. We also consider as the initial state the superposition $\sum_{m} c_m \psi_m (\mathbf{r})$ of wave functions of the form $\psi_m (\mathbf{r}) \propto (x +i y)^m e^{-(x^2 + y^2)/2}$.

Let us also comment here regarding the two-body contact interactions characterized by $g_0$. The existence and stability of vortex states in BECs with repulsive contact interactions, $g_0>0$, have been observed in numerous BEC experiments~\cite{Matthews1999, Madison2000, AboShaeer2001,Haljan2001,Hodby2001,Engels2002, Schweikhard2004,Leanhardt2002,Andersen2006,Beattie2013,Lin2009,Liu2024vortex}. In contrast, BECs with attractive contact interactions, $g_0<0$, are unstable in large atom numbers and consequently  collapse~\cite{Bradley1997}; hence, the creation of vortices in them is challenging. Nonetheless, it has been predicted that vortex states can also be stabilized in attractive BECs~\cite{Stringari1996,Clark2006}. However, this has not been achieved in any experiment yet. Therefore, we restrict ourselves in this work to the experimentally relevant case of a repulsive BEC.

\section{Superradiant threshold and its modification by the gauge field}
\label{app:sr_threshold}

To find the numerical threshold $\eta_c(\mathcal{B})$ for the SR phase transition, we solve Eqs.~\eqref{app_ss} for a given value of $\mathcal{B}$ on some range of $\eta$ starting from different initial conditions. We then compare the solutions and select the one that has the lowest total energy as explained above. The SR threshold $\eta_c(\mathcal{B})$ is passed when the condition $|\alpha|/\sqrt{N} \geq 0.01$ is satisfied. Similarly, we find the critical value of the synthetic magnetic field, $\mathcal{B}_c(\eta)$, for the single vortex nucleation.

To prove that the change of the superradiant threshold and the superradiant phase itself are due to a novel mechanism originating from the microscopic properties of the system we use an approach based on the perturbation theory~\cite{Zupancic2019P-Band}. For simplicity, we consider weakly interacting trap-free bosons inside the cavity such that as a first approximation we can ignore the two-body interactions. The Hamiltonian density describing the atom-cavity system takes the form,
\begin{equation} \label{eq:H-general}
    \hat{\mathcal H} = \hat{\mathcal H}_a  + \hat{\mathcal H}_c +  \hat{\mathcal H}_I, 
\end{equation}
where $\hat{\mathcal H}_a$ is a generic single-particle atomic Hamiltonian density, $\hat{\mathcal H}_c=-\Delta_c \hat{a}^\dag \hat{a}$ the bare cavity Hamiltonian, and $\hat{\mathcal H}_I$ describes the interaction between atoms and cavity photons
\begin{equation} \label{eq:HI}
    \hat{\mathcal H}_I =  \hbar U \hat{a}^\dag\hat{a} \hat {\mathcal B}
    +\hbar \eta (\hat{a}^\dag + \hat{a}) \hat\Theta.
\end{equation}
Here we have introduced order $\hat \Theta  =  \cos{(k_c \hat x)}$ and bunching $\hat {\mathcal B} =  \cos^2{(k_c \hat x)}$ parameters (operators). However, the term $\hbar U \hat{a}^\dag\hat{a} \hat {\mathcal B}$ has no fundamental contribution to the superradiant threshold and will be omitted.

To obtain a generic equation for the self-ordering (i.e., superradiant) threshold, we go to the mean-filed regime  $\langle\hat a\rangle = \alpha$. Near the critical point, $\alpha$ is very small such that we treat the interaction Hamiltonian~$\hat {\mathcal H}_I=\hbar \eta (\hat{a}^\dag + \hat{a}) \hat\Theta$ as a perturbation and obtain the following condition for the existence of a non-zero cavity field amplitude~\cite{Zupancic2019P-Band},
\begin{equation}
\label{eq:sr_crit}
\frac{4\hbar\eta^2 {\Delta}_{\mathrm{c}}}{{\Delta}_{\mathrm{c}}^2+\kappa^2} \chi  >1,
\end{equation}
where we have defined the single-particle response
function (i.e., static polarizability)
\begin{equation}
\label{eq:resp_func}
    \chi = 
\sum_n\frac{\Big|\left\langle\psi_n^{(0)}\right|\hat{\Theta}\left|\psi_0^{(0)}\right\rangle\Big|^2}{E_0^{(0)}-E_n^{(0)}}.
\end{equation}
Here $\psi_n^{(0)}$ are unperturbed eigenstates of the atomic Hamiltonian $\hat{\mathcal H}_a$ with eigenenergies $E_n^{(0)}$.

Equation~\eqref{eq:sr_crit} yields the superradiance threshold,
\begin{equation}
\label{eq:threshold-2nd-order}
\eta_c = \sqrt{\frac{{\Delta}_{\mathrm{c}}^2+\kappa^2}{4\hbar {\Delta}_{\mathrm{c}}\chi}}.
\end{equation}
Let us emphasize that the superradiant threshold~\eqref{eq:threshold-2nd-order} is rather general since we did not specify explicitly any atomic Hamiltonian $\hat{\mathcal H}_a$. The detail of the atomic Hamiltonian is instead encoded in the response function $\chi$, Eq.~\eqref{eq:resp_func}. We now elaborate this important point via two examples: 1) common self-ordering and 2) self-ordering under the influence of a gauge potential (i.e., our model).

\subsubsection{Common self-organization (in the absence of any gauge field)}

\begin{figure}[t!]
\centering
\includegraphics[width=0.48\textwidth]{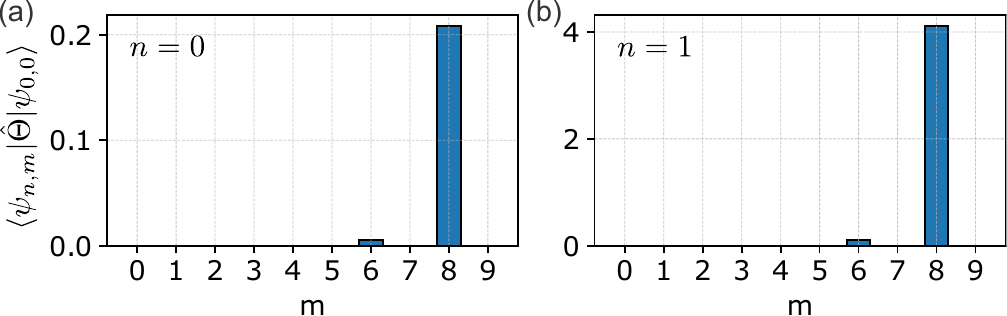}
\caption{Values of the matrix elements $\langle\psi_{n,m}^{(0)}|\hat{\Theta}|\psi_{0,0}^{(0)}\rangle$ for $n = 0$ (a) and $n = 1$ (b) with $m = 0,\ldots,9$. Note the different $y$-axis values in the two panels.}
\label{fig:matrix_elements}
\end{figure}

The single-particle atomic Hamiltonian for a uniform BEC contains only the kinetic-energy term, 
\begin{equation}
    \hat{\mathcal H}_a = \frac{\mathbf{\hat p}^2}{2M}.
\end{equation}
The eigenfunctions of this Hamiltonian are plane waves $\psi^{(0)}_k = e^{i\mathbf{k}\cdot \mathbf{r}}/\sqrt{2\pi}$, and the energies are given by $E^{(0)}_k = \hbar^2 k^2/2M$. Then the response function~\eqref{eq:resp_func} can be written as
\begin{equation}
    \chi = \sum_k\frac{\Big|\left\langle\psi^{(0)}_k\right|\hat{\Theta}\left|\psi^{(0)}_0\right\rangle\Big|^2}{E^{(0)}_0-E^{(0)}_k}.
\end{equation}
After simple integration, the response function yields
\begin{equation}
    \chi = -\frac{1}{2\hbar\omega_{\rm{rec}}},
\end{equation}
where $\omega_{\rm{rec}} = \hbar k_c^2/2M$. The transition to the superradiant phase then occurs at the critical pump amplitude,
\begin{equation} \label{eq:eta_c-common-SR}
\eta_c=\sqrt{\frac{\Delta_c^2+\kappa^2}{-2\Delta_c}}\sqrt{\omega_{\rm{rec}}}.
\end{equation}
This is the well-known result for the pump threshold for common self-ordering; see Ref.~\cite{Nagy2008}.

\subsubsection{Self-organization under the influence of a synthetic magnetic field} 

The free atomic BEC coupled to a synthetic gauge potential $\boldsymbol{\mathcal A}$ is described by the single-particle Hamiltonian, 
\begin{equation}
   \hat{\mathcal H}_a = \frac{(\mathbf{\hat p} - \boldsymbol{\mathcal{A}})^2}{2M}.
\end{equation}
The corresponding eigenenergies and wave functions are those of Landau levels and can be written in the symmetric gauge as
\begin{align} \label{eq:landau-levels}
    E_{n,m}^{(0)} &= \hbar\omega\Big(n + \frac{1}{2}\Big),\nonumber\\ 
    \psi_{n,m}^{(0)} &= \psi_{n,m}^{(0)}(r,\theta) = e^{im\theta}R_{n,m}(r),
\end{align}
where $n$ is the Landau level index ($n = 0,1,2,\ldots$), $m$ the angular momentum quantum number ($m=-n,-n + 1,-n + 2,\ldots$), and $\omega = \mathcal{B}/2M$. Here $(r,\theta)$ are polar coordinates and $R_{n,m}(r)$ is a radial function that involves generalized Laguerre polynomials~\cite{Ciftja2020}. Note that the energy spectrum is massively degenerate as $E_{n,m}^{(0)}$ does not depend on the quantum number $m$.

As a simple illustrative example, let us assume that the BEC is formed in the state $\psi_{0,0}^{(0)}$ (a non-vortex state). The matrix elements of the density order parameter $\langle\psi_{n,m}^{(0)}|\hat{\Theta}|\psi_{0,0}^{(0)}\rangle$ can then be calculated numerically. They are shown in Fig.~\ref{fig:matrix_elements} for $n = 0$ (a) and $n = 1$ (b) for various $m=0,\cdots,9$. Since the states $\psi_{0,m\neq0}^{(0)}$ are degenerate with the BEC ground state $\psi_{0,0}^{(0)}$ (i.e., $E_{0,m\neq0}^{(0)}-E_{0,0}^{(0)}=0$), any non-zero matrix element $\langle\psi_{0,m}^{(0)}|\hat{\Theta}|\psi_{0,0}^{(0)}\rangle$ leads to a divergent response function, Eq.~\eqref{eq:resp_func}. As a consequence the critical value of the pump strength $\eta_c$ reduces to zero; see Eq.~\eqref{eq:threshold-2nd-order}. 

Although in our full model the situation is more complicated due to the trapping potential and the two-body contact interactions, this simple example highlights that the presence of the artificial magnetic field significantly influences the superradiant threshold and superradiance itself in the microscopic level. In our full model with the trapping potential and the contact interactions, the eigensates of the system are no longer pure Landau levels and the BEC with vortices is a superposition of many $e^{im\theta}$-like states. Furthermore, many degeneracies are also likely lifted. Therefore, there is no divergence in the response function and the threshold does not fall to zero, rather only decreases---consistent with our numerical calculations.

\section{Photon scattering from a vortex as a density inhomogeneity}
\label{app:vortex_inhom}

In Fig.~\ref{fig:eta_c-B_c}(a), by increasing $\mathcal{B}$ slightly beyond $0.8M\omega_{\rm tr}$, a single vortex is nucleated in the center of the BEC density and the system simultaneously enters the SR regime. One might then conclude that the density inhomogeneity caused by the vortex plays the main role in the SR photon scattering. We address this important issue in this section and show that the photon scattering from the density inhomogeneity caused by a vortex sitting in the BEC center is not significant in general. 

\begin{figure}[t!]
\centering
\includegraphics[width=0.48\textwidth]{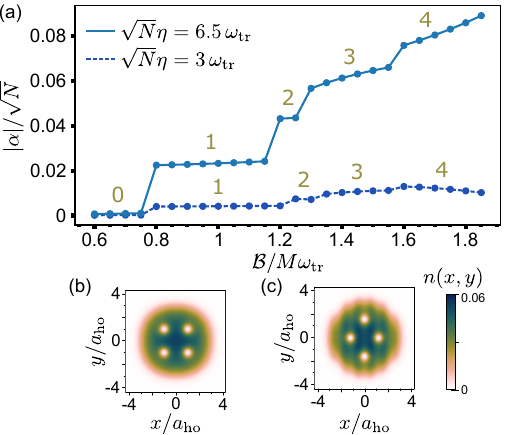}
\caption{Photon scattering from vortices. (a) The steady-state field amplitude $|\alpha|/\sqrt{N}$ as a function of the synthetic gauge field~$\mathcal{B}$ for different fixed pump strengths $\sqrt{N}\eta = 3\omega_{\rm tr}$ (dark blue dashed line) and $\sqrt{N}\eta = 6.5\omega_{\rm tr}$ (light blue solid line). The BEC density profile $n(\mathbf{r})$ is depicted for $\sqrt{N}\eta = 3\omega_{\rm tr}$ (b) and $\sqrt{N}\eta = 6.5\omega_{\rm tr}$ (c) at $\mathcal{B} = 1.8 M\omega_{\rm tr}$. The other parameters are the same as in Fig.~\ref{fig:eta_c-B_c} in the main text.}
\label{fig:SM_residual}
\end{figure}

To evaluate the contribution of the vortex to the photon scattering process we introduce the order parameter,
\begin{align}
    \Theta &= \langle \hat\Theta \rangle  = \int n(\mathbf{r})\cos(k_cx)\, dxdy \nn\\
    &= \int n_0(\mathbf{r})\cos(k_cx)\, dxdy - \int n_v(\mathbf{r})\cos(k_cx)\, dxdy \nn\\
    &= \Theta_0 - \Theta_v,
\end{align}
where we again decompose the total atomic density $n(\mathbf{r}) = n_0(\mathbf{r}) - n_v(\mathbf{r})$ into the vortex-free part $n_0(\mathbf{r})$ and the vortex density $n_v(\mathbf{r})$. The order parameter $\Theta$ asymptotically approaches one deep in the SR regime $\sqrt{N}\eta\to\infty$~\cite{Nagy2008}. We fix the pump strength to $\sqrt{N}\eta = 3\omega_{\rm tr}$ in the following. As a reference point, in the absence of the synthetic magnetic field  $\mathcal{B}=0$ (i.e., no vortex) the order parameter is close to zero, $\Theta=\Theta_0 \approx 2\times 10^{-5}$. On the other hand, for the synthetic magnetic field $\mathcal{B}= M\omega_{\rm tr}$ which stabilizes a single vortex in the center of the BEC, the order parameter takes the value $\Theta \approx -0.007$. This allows us to evaluate the input of the vortex to photon scattering as $\Theta_v = \Theta_0 -\Theta \approx -\Theta \approx 0.007$, implying that the density inhomogeneity caused by the vortex only minimally seeds the SR photon scattering. In other words, having a vortex which creates a density inhomogeneity does not necessarily guarantee SR photon scattering. This confirms that the rotational superradiance studied in this manuscript is \emph{not} a pure consequence of photon scattering from density inhomogeneities caused by vortices. Rather, it is a novel effect related to the fundamental changes of the system due to the broken TRS by the gauge field $\mathcal{B}$, as discussed in the previous appendix.

To further show this point, we also present in Fig.~\ref{fig:SM_residual}(a) the steady-state field amplitude $|\alpha|/\sqrt{N}$ as a function of the synthetic gauge field $\mathcal{B}$ for $\sqrt{N}\eta = 3\omega_{\rm tr}$ (the dark blue dashed curve). For comparison, we also include the behavior of the system for the pump strength $\sqrt{N}\eta = 6.5\omega_{\rm tr}$ (the light blue solid curve), which is the same as in Fig.~\ref{fig:phase-diagrams}(a) in the main text. By increasing the synthetic magnetic field $\mathcal{B}$ more vortices appear in the BEC (the indicated numbers in the plot). However, for the lower pump strength $\sqrt{N}\eta = 3\omega_{\rm tr}$ the number of photons scattered into the cavity mode stays almost below the SR condition ($|\alpha|/\sqrt{N} \geq 0.01$ is the numerical SR transition in our calculations as discussed in Appendix~\ref{app:sr_threshold}), even for the steady state containing four vortices. This again reinforces our conjecture of minimal photon scattering from density inhomogeneities caused by vortices. Let us also note that for $\sqrt{N}\eta = 3\omega_{\rm tr}$ the cavity-mediated interactions are insignificant compared to the strong pumping regime $\sqrt{N}\eta = 6.5\omega_{\rm tr}$, and the vortex-lattice structure is hence determined mainly by pairwise logarithmic interactions in the former case; cf.\ the BEC density profiles with four vortices for $\sqrt{N}\eta = 3\omega_{\rm tr}$ and $\sqrt{N}\eta = 6.5\omega_{\rm tr}$ in Fig.~\ref{fig:SM_residual}(b) and~(c). This, in turn, accounts for the decrease of the cavity field amplitude in the case of $\sqrt{N}\eta = 3\omega_{\rm tr}$ for the steady state containing four vortices by increasing $\mathcal{B}$, as the logarithmic interactions rearrange vortices to accommodate the forthcoming fifth vortex, hence reducing SR photon scattering. 

\section{Scaling behavior of the coherent cavity field as a function of the pump strength}
\label{app:scailing}

\begin{figure}[b!]
\centering
\includegraphics[width=0.48\textwidth]{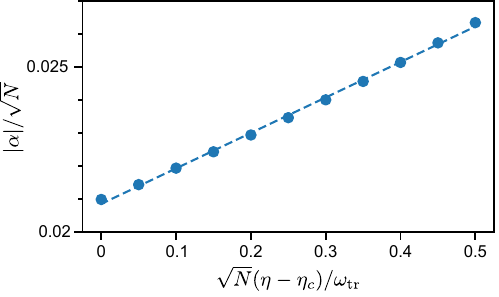}
\caption{Scaling behavior of the cavity field $\alpha/\sqrt{N}$ at the fixed synthetic magnetic field $\mathcal{B} = 1.5 M\omega_{\rm tr}$ as a function of the pump strength $\sqrt{N}\eta$ around the SR threshold $\sqrt{N}\eta_c \approx 4.5\omega_{\rm tr}$. Fitting the data numerically yields the relation $|\alpha(\mathcal{B} = 1.5 M\omega_{\rm tr})|/\sqrt{N}\propto\sqrt{N}(\eta-\eta_c)^\beta/\omega_{\rm tr}$, with the critical exponent $\beta=1$ for our finite-size system.}
\label{fig:SM_scailing}
\end{figure}

In this section, we explore numerically the scaling behavior of the coherent cavity field $|\alpha|/\sqrt{N}$ at the fixed synthetic gauge field $\mathcal{B} = 1.5M\omega_{\rm tr}$ as a function of the pump strength $\sqrt{N}\eta$ in the vicinity of the SR phase transition $\sqrt{N}\eta_c \approx 4.5\omega_{\rm tr}$. To this end, we fit the data in the pump interval $\sqrt{N}\eta\in [4.5\omega_{\rm tr},5\omega_{\rm tr}]$ as shown in Fig.~\ref{fig:SM_scailing} and obtain the scaling relation $|\alpha(\mathcal{B} = 1.5 M\omega_{\rm tr})|/\sqrt{N}\propto\sqrt{N}(\eta-\eta_c)^\beta/\omega_{\rm tr}$ for our finite-size system, with the critical exponent $\beta=1$. Ignoring possible finite-size effects, this is in contrast to the common SR scaling $|\alpha(\mathcal{B} = 0)|/\sqrt{N}\propto\sqrt{N}(\eta-\eta_c)^{1/2}/\omega_{\rm tr}$ where the critical exponent is $1/2$.



%

\end{document}